# Fabrication of Metasurfaces on Building Construction Materials for Potential Electromagnetic Applications in the Microwave Band

Zacharias Viskadourakis [1,*], Konstantinos Grammatikakis [2], Klytaimnistra Katsara [1,3], Argyri Drymiskianaki [2] and George Kenanakis [1]

[1] Institute of Electronic Structure and Laser (IESL), Foundation for Research and Technology—Hellas (FORTH), N. Plastira 100, Vassilika Vouton, 70013 Heraklion, Greece; klyto.katsara@iesl.forth.gr (K.K.); gkenanak@iesl.forth.gr (G.K.)
[2] Materials Science and Technology Department, University of Crete, Vassilika Vouton, 70013 Heraklion, Greece; mst1642@edu.materials.uoc.gr (K.G.); adrym@materials.uoc.gr (A.D.)
[3] Department of Agriculture, Hellenic Mediterranean University—Hellas, Estavromenos, GR-714105 Heraklion, Greece
* Correspondence: zach@iesl.forth.gr; Tel.: +30-2810-391921

**Abstract:** Energy self-sufficiency, as well as optimal management of power in buildings is gaining importance, while obtaining power from traditional fossil energy sources is becoming more and more expensive. In this context, millimeter-scale metasurfaces can be employed to harvest energy from microwave sources. They can also be used as sensors in the microwave regime for efficient power management solutions. In the current study, a simple spray printing method is proposed to develop metasurfaces in construction materials, i.e., plasterboard and wood. Such materials are used in the interior design of buildings; therefore, the implementation of metasurfaces in large areas, such as walls, doors and floors, is realized. The fabricated metasurfaces were characterized regarding their electromagnetic performance. It is hereby shown that the investigated metasurfaces exhibit an efficient electromagnetic response in the frequency range (4–7 GHz), depending on the MS. Thus, spray-printed metasurfaces integrated on construction materials can potentially be used for electromagnetic applications, for buildings' power self-efficiency and management.

## 1. Introduction

Metamaterials (and their 2-dimensional counterparts called metasurfaces (MSs)) can be defined as man-made materials that exhibit extraordinary electromagnetic properties not found in natural materials. Such properties include perfect absorption, large positive refractive index, negative refractive index, magnetism at optical frequencies, zero reflection through impedance matching, etc. [1–6]. Therefore, they can be used in a great variety of applications (i.e., electromagnetic shields, sensors, splitters, isolators, harvesters, modulators, etc.), in a wide frequency range [7–14]. The building block of a metasurface (MS) is called the meta-atom and it is of specific shape and dimensions. The electromagnetic (EM) properties of MSs are directly related to those two parameters. It also depends on the distribution of the meta-atoms in the space. Hence, by tuning the size, the shape and spatial distribution of the meta-atoms, the EM properties of MSs can be gently manipulated for optimal performance in a specific regime of the electromagnetic spectrum. Especially for EM applications in the microwave regime, the size of the corresponding meta-atoms should be of the order of several millimeters.

An application of metamaterials, which has recently gained considerable interest, is their use in the construction of smart buildings—i.e., buildings in which (metamaterial) technology is employed to enable the efficient use of resources in a safe and comfortable environment for the occupants. In particular, seismic metamaterials are employed to enhance the antiseismic behavior of buildings [15–17]. Moreover, acoustic metamaterials are used to transform acoustic waves and low frequency vibrations into electrical power [18–21], while metamaterials performing invisible frequencies are used to enhance the efficiency of solar cells [22–24]. Additionally, metamaterial sensors are used to detect thermal fluctuations and changes in order to optimize temperature control in a building [25–27]. Thus, modern buildings (hospitals, airports, universities, train stations, factories, trade centers, skyscrapers, etc., all of which need tremendous amounts of power) include novel technologies in order to harvest energy from various sources, such as heat, electromagnetic signals, solar energy, etc. Furthermore, modern buildings incorporate innovative solutions for power management in order to reduce energy consumption. Such technological improvements are actually based on both sensitive sensors and efficient harvesters [28–32].

Recently, millimeter-scale MSs have been developed for potential use either as sensors or as harvesters in the microwave regime. More specifically, 3-dimensional (3D) printed Split Ring Resonators (SRRs) exhibit significant EM response to slight dielectric constant changes, enabling their potential use for sensing applications in the microwave regime [2]. In addition, 3D printed, stand-alone, cut-wire MSs were developed to collect microwave energy at frequencies around 2.4 GHz and transform it into electrical power. With an efficiency of ~7%, such MSs are quite promising for potential energy harvesting applications in the mm-wavelength band [33]. Considering that microwave sources (wi-fi hubs, microwave ovens, antennas, electronic circuits, etc.) can be present inside as well outside a building, the use of such MSs for buildings' power management and energy conservation is a potentially promising approach. Notably, so far, there have been no reports on the use of microwave MSs in smart buildings—neither for energy harvesting nor for sensing. Thus, their application towards buildings' self-efficiency and optimal power management would be advantageous.

Taking into account all the above, in the current study, we propose the fabrication of microwave-efficient MSs in construction materials, namely, wood and plasterboard, employing the spray deposition method. Both plasterboard and wood are basic components for modern buildings' interior design. Interior walls and doors, as well as floors and ceilings, made of plasterboard and wood, make up a large in-house area, which is appropriate for integrating sensors and harvesters to capture microwave signals from surroundings. The captured EM signal could either be used for transformation into electrical power (i.e., through the cut-wire MSs) or be recorded (i.e., through SRRs), and provide useful information for the optimization of power management for buildings' power self-efficiency. Crucial preconditions for such integration include the simple MS fabrication technique, which is compatible with building construction routes. In this context, spray deposition is the most favorable method, since it is highly compatible with normal painting techniques used in building construction. In addition, the use of water-based eco-friendly paints suits the spray technology.

The spray-deposited MSs were characterized regarding their morphology structure and electrical properties. Moreover, their EM performance was also studied and compared to traditionally grown MSs—such as Printed Circuit Board ones (PCB-printed), which are considered as reference MSs—in this study. It is shown that spray-printed MSs show distinct EM characteristics that are qualitatively similar to the PCB-printed MSs. The simple growth method, as well as the efficient EM performance of the grown MSs, enables their quick and easy development onto large surfaces, such as building walls, and empowers their potential use in applications directed at buildings' energy self-efficiency.

## 2. Materials and Methods

### 2.1. Metasurface Fabrication

The investigated MSs were grown using the conventional spray deposition technique. Three different precursors were used as painting materials. The first precursor is homemade and consists of water in which graphite fillers are homogeneously dispersed.

Specifically, 20 mL of distilled water was put in a beaker along with 1 g of graphite. The solution was sonicated for 30 min, and then it was left for 10 min. The sonication process was repeated 3 times so that a total sonication time of 2 h was applied. Hence, a final dispersion suitable for the spray process was achieved.

The second precursor is a commercially available conductive paint. It is a water soluble, non-toxic, water-based dispersion of carbon pigment in natural resin supplied by Bare Conductive (Bare conductive LTD, London, UK). The paint possesses high conductivity, enabling its use to develop/repair electronic circuits. However, due to its high viscosity, it cannot be used in its current form. Thus, a portion of the paste was diluted into distilled water in a weight ratio of 9/1 (it must be noted that other mass ratios were also tried for both precursors. Nevertheless, high mass ratios result in increased viscosity of the precursor, leading to issues regarding airbrush nozzle blocking and poor spray plume coming out of the nozzle). The third precursor is the commercially available liquid YShield HSF54 (YShield GmbH $ Co. KG, Ruhstorf, Germany), used for EM shielding applications. It basically consists of water in which carbon nanofillers have been uniformly dispersed in a weight ratio of 15%. This water-based pigment was used in its current form without any further treatment.

An appropriate amount of a precursor was inserted into a standard airbrush that is used for painting jobs (Figure 1a). The airbrush was supplied by air pressure (~3 bar) through an oil-free air compressor. The distance between the airbrush and the printing surface was kept at ~25 cm. The whole process took place at room temperature and ambient atmosphere.

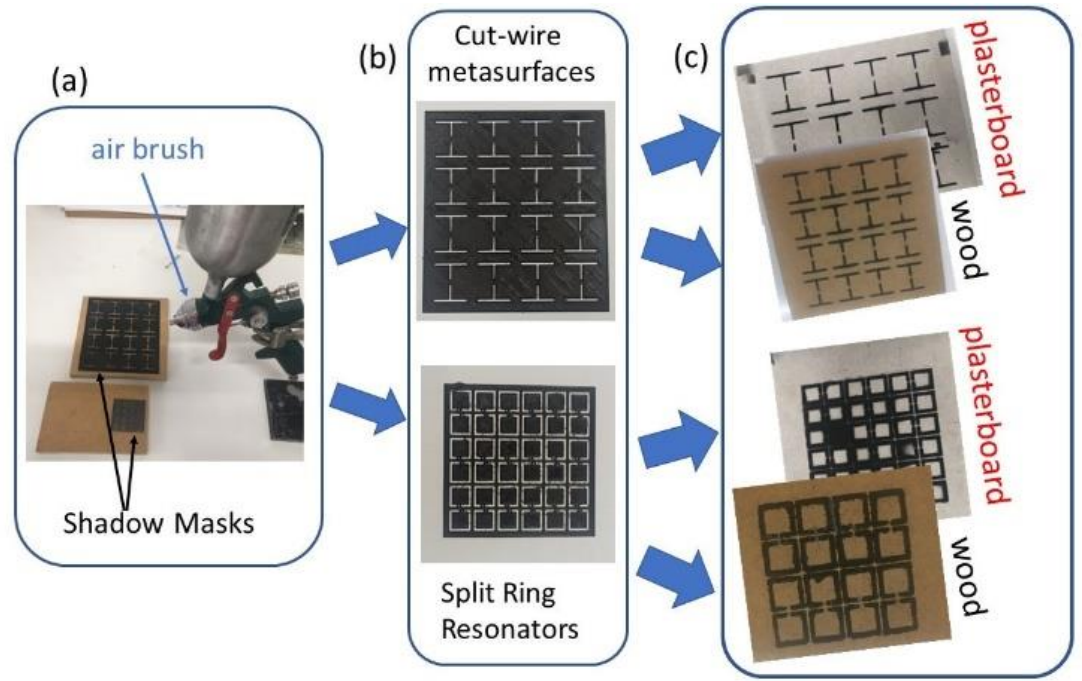

**Figure 1.** (**a**) Experimental set-up for the spray-printing fabrication of the investigated metasurfaces (MSs). The airbrush as well as the shadow masks are shown. (**b**) Details regarding the shadow masks used for the spray deposition process. (**c**) Typical view of spray deposited MSs on wood and plasterboard, using both the cut-wire and Split Ring Resonator (SRR) shadow masks.

In order to achieve MSs of the desired shape, dimensions and spatial distribution, appropriate shadow masks were built, employing the so-called Fused Deposition Modeling (FDM) method, which is a conventional additive manufacturing process for developing complex 3D structures. Polylactic acid (PLA) filament has been used as the starting material. Notably, the FDM method, along with PLA filaments, has been previously used to develop stand-alone MSs for sensing and energy harvesting applications in the microwave regime [2,33]. Therefore, it is a perfect combination for the formation of shadow masks in the mm range. Two different shadow masks were constructed: one with cut-wire MSs and another with SRRs, as shown in Figure 1b. Both include several MS units, and each unit has dimensions given in Table 1.

**Table 1.** Shadow mask drawings and dimensions.

| Metasurface Type | Drawing | Dimensions |
| --- | --- | --- |
| Cut-wire | 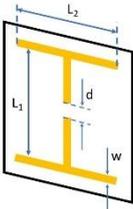 | $L_1$ = 23.1 mm<br>$L_2$ = 22.8 mm<br>w = 1.6 mm<br>d = 3 mm |
| Split Ring Resonator | 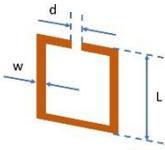 | L = 10 mm<br>w = 1 mm<br>d = 0.4 mm |

The shadow masks were attached at the top of the desired surface and then the precursor was deposited, employing the spray method, as described above. Then, the surfaces were left at room temperature for 30 min so that the printed MSs dried. Both printing and dry procedures were repeated several times so that the MSs became thicker. Following the above-described procedure, several different samples were constructed (Figure 1c).

*2.2. Characterization of the Metasurfaces*

The deposited MSs were characterized regarding their shape and dimensions through optical microscopy experiments. To this end, an AP-8 microscope (Euromex Microscopen bv., Arnhem, The Netherlands) was used, achieving magnifications up to ×80. Moreover, the structural properties of MSs were studied through X-ray Diffraction (XRD) experiments, using a Rigaku (RINT 2000) Cu $K\alpha$ X-ray diffractometer. Furthermore, Raman experiments were performed at room temperature using a modified LabRAM HR Raman Spectrometer (HORIBA Scientific, Kyoto, Japan), where the microscope was coupled with a 50× microscopic objective lens that delivered the excitation light and collected the Raman signal. Raman excitation was achieved with a 532 nm central wavelength solid state laser module (maximum laser output power of 90 mW). The maximum power on the sample surface was 4 or 9 mW, depending on the laser intensity applied (10% or 25% laser intensity, respectively) after the use of a neutral density filter of 5% transmittance. The Raman spectral range was set to be from 300 to 1800 cm$^{-1}$. Acquisition time for each measurement was 5 s and with 3 accumulations in each sample's point. Finally, a 600 groves/mm grating was used, resulting in a Raman spectral resolution of around 2 cm$^{-1}$.

In addition, FT-IR experiments were carried out, using a Bruker Vertex 70v FT-IR vacuum spectrometer (Bruker, Billerica, MA, USA) equipped with an A225/Q Platinum ATR unit with single reflection diamond crystal, which allows the infrared analysis of unevenly shaped solid samples through total reflection measurements in a spectral range of 4000–350 cm$^{-1}$. Interferograms were collected at 4 cm$^{-1}$ resolution (8 scans), apodised with a Blackman–Harris function and Fourier transformed with two levels of zero filling to yield spectra encoded at 2 cm$^{-1}$ intervals. Before scanning the samples, a background diamond crystal was recorded, and each sample spectrum was obtained by automatic subtraction. For each measurement, the samples were carefully placed under the ATR press, while after every measurement the sample area and the tip of the A225/Q ATR unit were cleaned with pure ethanol (Et-OH; Sigma-Aldrich, Munich, Germany).

The resistance MSs was also measured, employing a conventional 2-probe technique (Figure 2a). Specifically, metallic needle-like contacts were attached at the ends of metallic lines of the MSs. The needles were connected to a sensitive multimeter (SDM 3065X, Siglent.eu JR Special Electronics, Helmond, The Netherlands) so that the resistance of specific elements of each MS (i.e., see Figure 2a) was directly measured. Experimental data were obtained for all MSs printed onto a surface, and corresponding statistical analysis was performed.

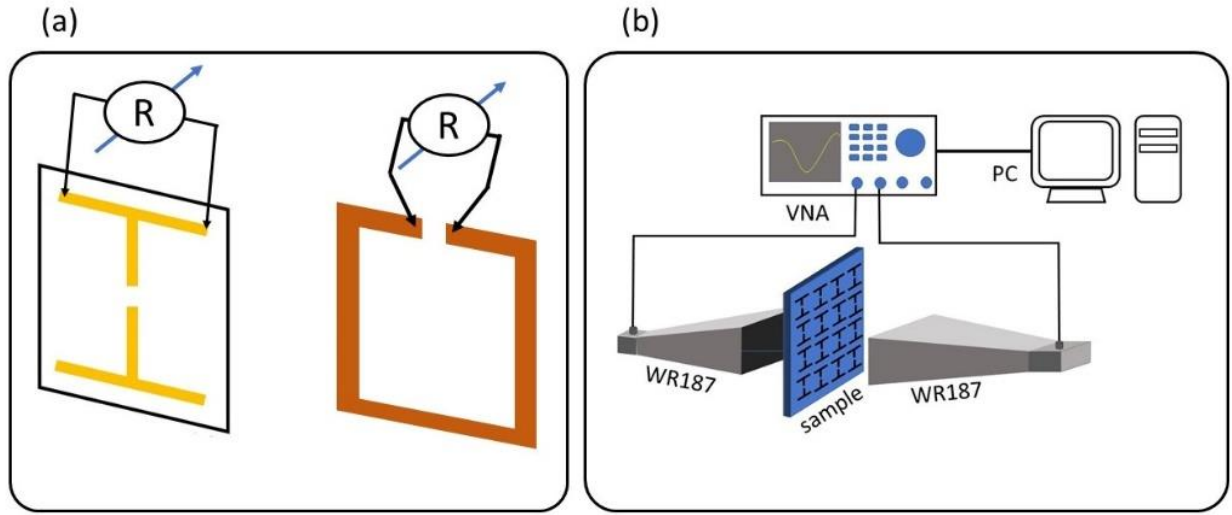

**Figure 2.** (**a**) Experimental set-up for measuring the resistance of MSs. (**b**) Experimental set-up for measuring the electromagnetic response of spray printed MSs.

The EM response of the MSs was measured through transmission measurements. In this context, a Vector Network Analyzer (VNA) was used (P9372A Streamline Vector Network Analyzer, Keysight, CA, USA) in combination with WR187 standard horn antennas (Figure 2b). Details regarding the technique and measurement performance have been previously described [2,33,34]. Corresponding EM measurements for PCB-printed MSs are also presented for direct comparison.

## 3. Results and Discussion

### 3.1. Structural and Morphological Characterization

Figure 3 shows typical optical microscopy images of the spray printed MSs, i.e., cutwire MSs on plasterboard (Figure 3a), and SRRs printed on wood (Figure 3b). Similar pictures were obtained for all MSs and are listed in Table 2. All MSs printed on plasterboard show similar characteristics. As shown in Figure 3a, printed MSs exhibit good geometrical features. Specifically, printed lines are well-defined and corners and T-crosses are sharp enough, while gaps are well shaped, exhibiting negligible roundness.

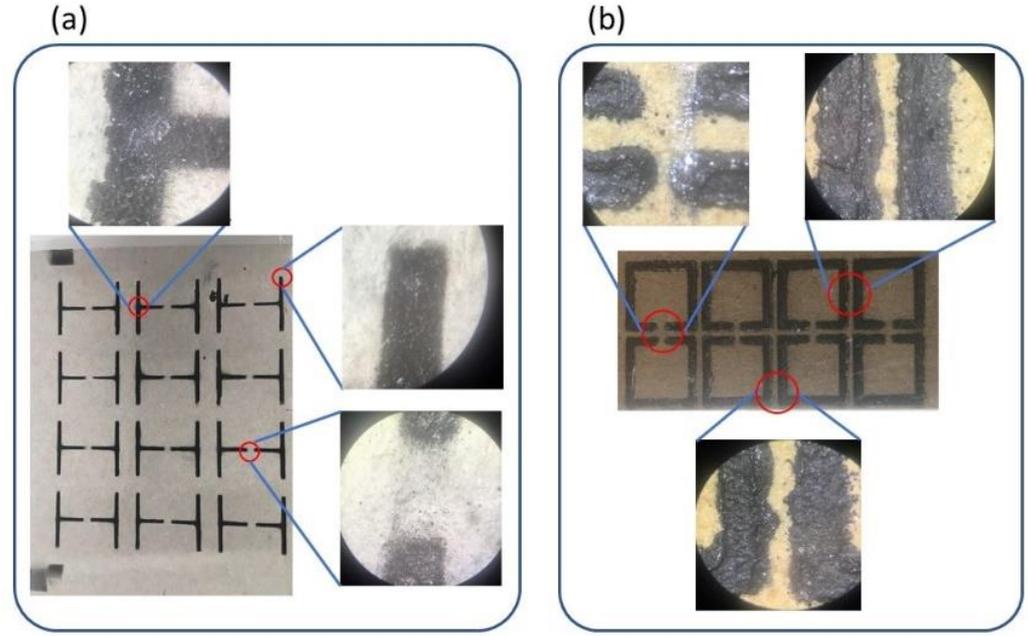

**Figure 3.** Optical microscopy pictures of printed MSs. (**a**) Details regarding the shape of (**a**) cut-wire MS, printed on plasterboard, and (**b**) SRRs printed on wood.

**Table 2.** Types of MSs printed, and their corresponding measured dimensions.

| Printing Surface | MS Type | Precursor | Dimensions (mm) | | | |
|---|---|---|---|---|---|---|
| Plasterboard | Cut-wire | Graphite | $L_1$ = 23.0 ± 0.1 | $L_2$ = 22.7 ± 0.2 | D = 2.7 ± 0.3 | w = 1.9 ± 0.3 |
| | | Carbon paste | $L_1$ = 23.0 ± 0.1 | $L_2$ = 22.6 ± 0.1 | d = 2.9 ± 0.1 | w = 1.57 ± 0.03 |
| | | HSF54 | $L_1$ = 23.2 ± 0.2 | $L_2$ = 22.8 ± 0.2 | d = 2.9 ± 0.1 | w = 1.7 ± 0.2 |
| | SRR | Graphite | L = 10.5 ± 0.4 | w = 1.09 ± 0.07 | d = 0.6 ± 0.1 | |
| | | Carbon paste | L = 10.2 ± 0.4 | w = 1.07 ± 0.07 | d = 0.6 ± 0.1 | |
| | | HSF54 | L = 10.4 ± 0.4 | w = 1.08 ± 0.07 | d = 0.6 ± 0.1 | |
| Wood | Cut-wire | Graphite | $L_1$ = 23.0 ± 0.1 | $L_2$ = 22.7 ± 0.2 | d = 2.9 ± 0.1 | w = 1.8 ± 0.1 |
| | | Carbon paste | $L_1$ = 23.3 ± 0.1 | $L_2$ = 22.5 ± 0.2 | d = 2.7 ± 0.1 | w = 1.8 ± 0.1 |
| | | HSF54 | $L_1$ = 23.3 ± 0.1 | $L_2$ = 22.8 ± 0.2 | d = 2.7 ± 0.1 | w = 1.8 ± 0.1 |
| | SRR | Graphite | L = 10.6 ± 0.4 | w = 1.08 ± 0.07 | d = 0.6 ± 0.1 | |
| | | Carbon paste | L = 10.3 ± 0.4 | w = 1.07 ± 0.07 | d = 0.6 ± 0.1 | |
| | | HSF54 | L = 10.5 ± 0.4 | w = 1.08 ± 0.07 | d = 0.6 ± 0.1 | |

In contrast, MSs printed in wood display inferior shape features (Figure 3b). Printed lines are not well-shaped, the distance between neighbor lines varies due to the poor shape of the lines, and corners show roundness, while gaps differ from each other. Such an inferior printing result in wood could possibly be attributed to the surface itself. Wood surface is much rougher than plasterboard. In general, rough surfaces do not favor the deposition of uniformly shaped MSs. Notably, since the EM performance of the MSs is closely related to their shape, such discrepancies in the shape of the printed MSs should affect their EM response. In other words, plasterboard-printed MSs are expected to perform better than the wood-deposited ones (this will be shown later).

Regarding the dimensions of the printed MSs, results are presented in Table 2. Measured values are consistent with dimensions of the shadow masks; thus, the printed MSs possess rather good dimensions.

Figure 4a shows typical XRD spectra of MSs on plasterboard. Spectra of both carbon paste and HSF54 precursors are shown. Regardless of the precursor, peaks indexed to graphite structure [35,36] are observed (black stars). All other peaks are attributed to the plasterboard surface. Notably, plasterboards are composed of several different materials, such as gypsum, paper, fibers (wool and/or glass), plasticizers, foaming agents, surfactants and other additives. Therefore, the complex XRD spectrum, presented in Figure 4a, can be plausibly attributed to the composition of the plasterboard. Similar spectra were obtained for MSs grown on wood, as well as for MSs printed using a graphite precursor. Moreover, Figure 4b shows typical FT-IR spectra of MSs deposited with all precursors (plasterboard). The FT-IR spectrum of bare plasterboard is also included for comparison. All peaks observed can be identified as FT-IR peaks corresponding to carbon, graphite oxide and gypsum, according to the literature [37–39] . Moreover, the noise in the range of 2000–2500 cm$^{-1}$ is most likely attributable to water [40]. Plasterboard easily absorbs humidity, which could be inspected in the FT-IR spectrum. Notably, similar FT-IR spectra are obtained for MSs printed on wood surfaces. In addition, Figure 4c shows the Raman spectra of MSs deposited on plasterboard. Two major peaks are observed at ~1350 cm$^{-1}$ and ~1590 cm$^{-1}$, respectively, for all printed MSs. The latter peak (G band) is associated with the graphite structure where carbon atoms exhibit a sp$_2$ hybridization [41,42]. The former (D band) is related to the defects included in the lattice structure of carbon materials [41].

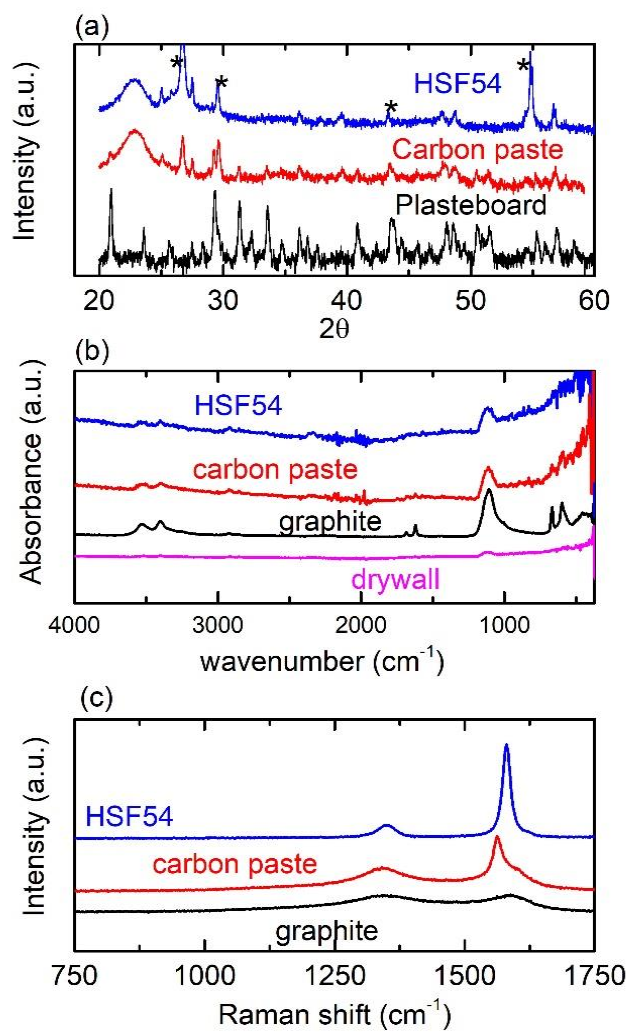

**Figure 4.** Structural and molecular characterization of MSs, printed on plasterboard, for various precursors, i.e., (**a**) XRD spectra. * correspond to carbon nanoinclusions. (**b**) FT–IR spectra and (**c**) Raman spectra.

*3.2. Resistance of the MSs*

Figure 5 shows the resistance measurement distribution for all MSs printed on plasterboard (both cut-wire and SRRs). Corresponding statistical analysis is presented in Table 3. Regardless of the precursor, there are a few similarities observed in all histograms. Specifically, the distribution of the measurements is rather wide in all cases. Moreover, all distributions are asymmetric, with the data population to be shifted to values lower than the average (i.e., see Table 3). Moreover, the calculated statistical error is ~30% in all cases. The above-mentioned evidence clearly depicts the inhomogeneous conducting behavior of the MSs. Such resistance diversity likely suggests that the fabricated MSs are not the same, corroborating the diversities observed by optical microscopy. Furthermore, spray deposition could possibly result in MSs of various thicknesses. The precursor passes through the airbrush nozzle at high pressures, making a cone of droplets moving towards the printing surface. Droplet distribution, of course, is highly inhomogeneous; therefore, it is likely to lead to the development of MSs with various thicknesses, which could further result in the variation of resistance.

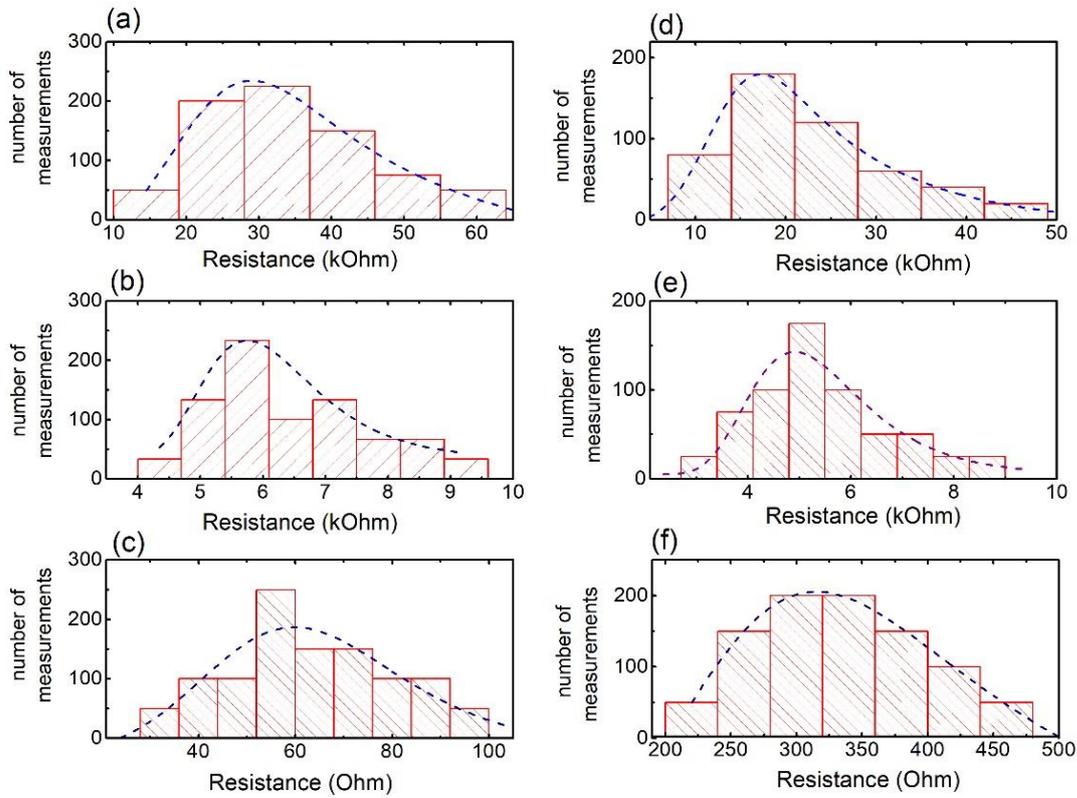

**Figure 5.** Resistance measurement of MSs printed on plasterboard. Cut–wire MS, made of (**a**) graphite, (**b**) carbon paste and (**c**) HSF54. SRRs deposited, using (**d**) graphite, (**e**) carbon paste and (**f**) HSF54, respectively. Dash lines are guides to the eye.

**Table 3.** Statistical analysis of resistance for various MSs deposited on plasterboard.

| Metasurface Type | Precursor | Average Resistance (Ohm) | % Population below Average Resistance |
|---|---|---|---|
| Cut-wire | Graphite | $(35 \pm 11) \times 10^3$ | 63.0 |
|  | Carbon paste | $(6 \pm 2) \times 10^3$ | 62.5 |
|  | HSF54 | $63 \pm 17$ | 62.0 |
| Single Ring Resonator | Graphite | $(22 \pm 10) \times 10^3$ | 76 |
|  | Carbon paste | $(6 \pm 2) \times 10^3$ | 60 |
|  | HSF54 | $335 \pm 60$ | 67 |

Comparing among different precursors, it is obvious that the graphite-printed MSs show the highest resistance, lying in the range of a few tens of kOhms (Figure 5a,d). In addition, carbon paste MSs possess resistance values in the range of a few kOhms (Figure

5b,e)—i.e., ~5–6 times lower. On the other hand, the HSF54 MSs exhibit resistance values of a few tens of Ohms (Figure 5c,f). Such a large difference can be attributed to the precursors used. More specifically, the HSF54 precursor (commercially available) is dedicated to EM shielding applications, and thus it is developed to exhibit high conductivity (low resistance). It is important to note that, the lower their resistance, the better EM performance MSs exhibit [2]. Thus, the HSF54 MSs are expected to perform better than the others regarding their EM response.

Figure 5d–f show the resistance of the SRRs deposited on the plasterboard. Results derived from corresponding statistical analysis are also presented in Table 3. Resistance of SRRs qualitatively shows similar behavior to the cut-wire MSs. Thus, there is large diversity in resistance values, which may be attributed to the deposition process, as previously described. Moreover, resistance distribution is asymmetric, with the value population to be shifted towards values lower than the average. Additionally, statistical error is rather large. All these observations suggest large differences among neighbor SRRs, probably coming from the deposition process itself. Additionally, the HSF54 SRRs (Figure 5f) exhibit the lowest resistance among all printed MSs, while the graphite-printed SRRs show the largest resistance (Figure 5d). Therefore, the latter MSs are expected to show the poorest EM behavior, while the former are expected to show better EM performance.

Here, it must be noted that qualitatively similar results are obtained for MSs printed on a wood surface (i.e., see Supplementary part Figure S1). Nevertheless, the resistance values are greater, especially when using graphite and carbon paste precursors. The poor conducting behavior can be attributed to the wood surface, which seems to be more absorbent than the plasterboard. Thus, the deposited precursor is absorbed, preventing the graphite/carbon nanoinclusions from developing a bulk continuous path, thereby suppressing the charge movement among fillers and the consequent resistance increment. In contrast, the HSF54 MSs (both the cut-wire and the SRRs) show good conductivity characteristics—i.e., low resistance—enabling good EM performance.

*3.3. Electromagnetic Response of the MSs*

Figure 6a shows the EM response of all spray-printed SRRs on plasterboard. The orientation of the incident EM wave with respect to the orientation of the SRRs is shown in the inset picture (TE polarization). Distinct differences among SRRs using different precursors are observed. Specifically, the poorest EM performance is exhibited by graphite SRRs, showing negligible absorption. Carbon paste MSs, on the other hand, show moderate absorption (~5 dB) in the whole measure frequency range, whereas HSF54 SRRs demonstrate the best EM performance among all printed SRRs, reaching ~ −10 dB above 6 GHz. They also show a shallow but well-defined minimum at ~6.6 GHz, which is indicative of resonance. Notably, the PCB-printed SRRs resonate at ~6.7 GHz, corroborating the resonance frequency of HSF54 samples, indicating good EM performance of the latter. Moreover, the EM behavior of the HSF54 MSs is qualitatively similar to that of the stand-alone, 3D-printed SRRs that were previously studied [2], signifying their respectable performance. The fact that HSF54 performs the best among all spray-printed SRR, is directly related to its resistance, as previously discussed. Graphite SRRs exhibit the highest resistance of all SRRs, while they also show the poorest EM response. On the other hand, carbon paste SRRs are better than the graphite ones since they have lower resistance. However, none of them show any indication of resonance. On the contrary, HSF54 MSs, having the lowest resistance (i.e., a few decades of Ohms) of the three, show not only the greatest EMI response, but they also resonate at a certain frequency.

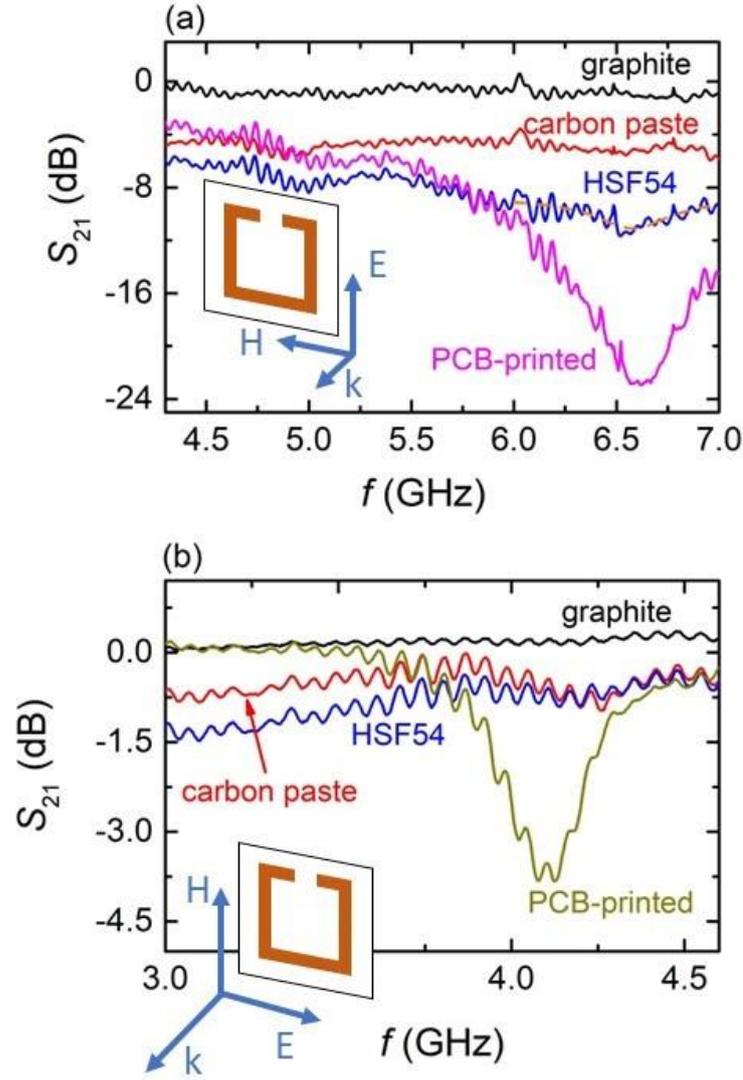

**Figure 6.** (**a**). $S_{21}$ spectra of all SRR MSs printed on plasterboard (TE polarization). The electromagnetic response of the PCB–printed MS is also presented for comparison. The orange dash line is guide to the eye. Incident electromagnetic wave, with respect to the SRR orientation is shown in the inset drawing. (**b**). $S_{21}$ spectra of all SRRs printed on plasterboard (TM polarization). The electromagnetic response of the PCB–printed MS is also presented for comparison. Direction of the electromagnetic wave, with respect to the SRR orientation is shown in the inset drawing.

Figure 6b shows the EM performance of all spray-printed SRRs in the TM polarization (i.e., see Figure 6b inset). All SRRs show rather low absorption in the whole measurement range. Moreover, none of them resonates, in contrast to the PCB–printed SRRs, which show resonance at 4.2 GHz. Nevertheless, there is a slight difference among all three spray-printed SRRs regarding their EM performance. Specifically, the graphite MSs do not absorb (i.e., $S_{21} \sim 0$), while carbon paste SRRs show $S_{21} \sim -0.5$ dB. In addition, the HSF54 MSs depict $S_{21}$ values of $\sim -1$ dB. Thus, compared to all three SRRs, the HSF54 ones perform the best of all, even though they do not resonate.

Here, it must be noted that ripples are shown in all $S_{21}$ spectra. This is an artifact revealed due to the fact that the whole experimental set-up (Figure 2b) is placed in an open space area, without using the anechoic chamber. Thus, the incident wave is transmitted in the air and the signal detected by the receiving antenna most likely includes parasitic contributions coming from electronic devices located near the experimental set-up. However, the ripples' width is small enough that the main features of the EM response of the studied MSs can be intrinsically revealed without any assumptions. This gives further credence to the investigated MSs since the exhibited performance shows up in real-life conditions, enhancing their potential applications for energy self-efficiency and power management in smart buildings.

The EM response of all cut-wire MSs on plasterboard is shown in Figure 7a. The curves were obtained in the $E_y$ polarization (i.e., Figure 7b, upper drawing). It is seen that both graphite and carbon paste-made MSs show zero $S_{21}$ values in the whole measured frequency range, which is indicative of eliminated EM response. The HSF54 MS, however, exhibits a broad, though well-defined, minimum at ~4 GHz, suggestive of resonance. The $S_{21}$ minimum is shallow (i.e., less than −4 dB) and noisy; nevertheless, it occurs at the same frequency where the PCB-printed cut-wire MS resonates, suggesting its intrinsic nature. Similar resonance results have been obtained for 3D-printed cut-wire MSs as well [33]. Moreover, by turning the MS 90° with respect to electric field counterpart of the incident EM wave, the $S_{21}$ spectrum, in the $E_x$ polarization, is obtained (Figure 7b). A slight shift (~100 MHz) towards higher frequencies is observed with respect to the $E_y$ polarization. Such a shift has also been obtained for PCB-printed MSs (supplementary part Figure S2); thus, the EM behavior of the HSF54 spray-printed MSs qualitatively resembles the behavior of the PCB-printed ones.

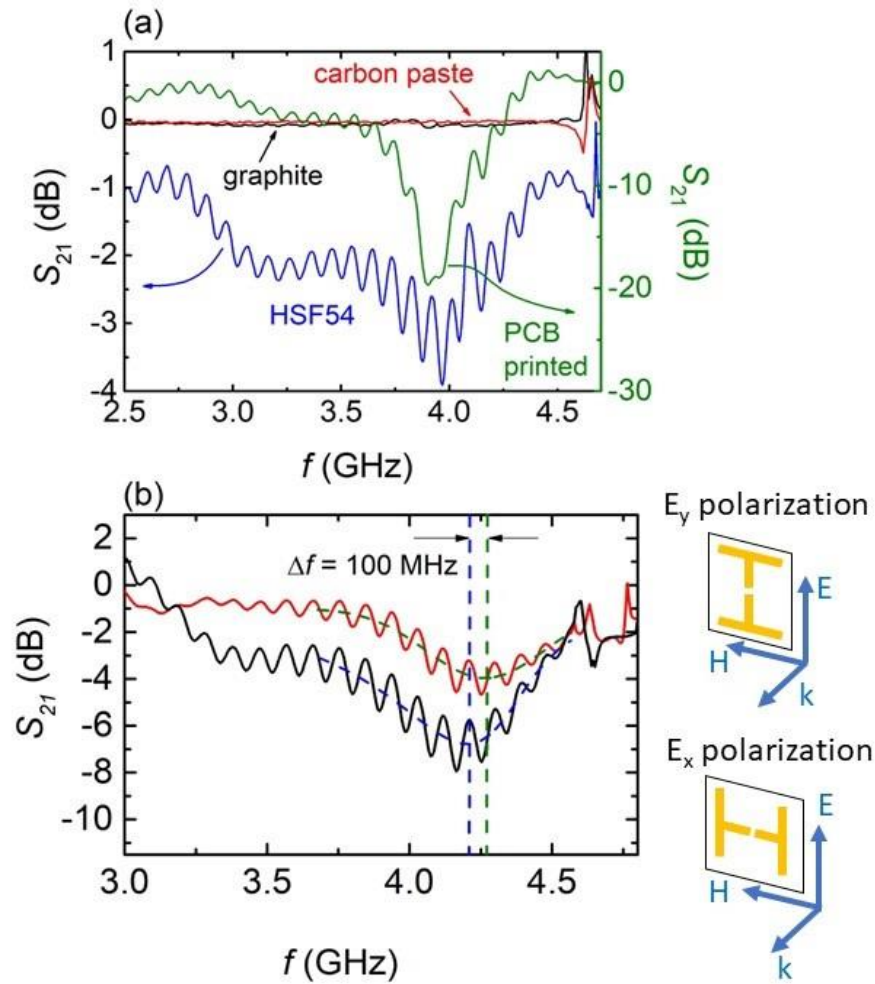

**Figure 7.** (**a**). $S_{21}$ spectra of all cut–wire MSs printed on plasterboard. The EM response of the PCB-printed MS is also presented (green solid line), for comparison. The spectra were obtained in the $E_y$ polarization as shown in the next panel (**b**). $S_{21}$ spectra for the cut–wire MS, made of HSF54 precursor, for different polarizations. The upper drawing corresponds to the polarization, used for obtaining the black curve ($E_y$ polarization), while the lower drawing shows the orientation of the MS, so as the red curve was obtained ($E_x$ polarization). The dash lines are guides to the eye.

In addition, Figure 8 shows the EM performance for all printed cut-wire MSs on wood for both $E_y$ (Figure 8a) and $E_y$ (Figure 8b) polarizations. The HSF54 MSs exhibit a broad minimum at frequencies around 4.5 GHz, regardless of the polarization. Although broad, the minimum is well-defined and in the $E_x$ polarization reaches values as low as −10 dB. Moreover, compared to those printed on plasterboard, the wood-printed MSs behave in a similar manner. On the other hand, neither graphite nor carbon paste cut-wire MSs show any feature of resonance. Even more, SRRs printed on wood do not exhibit any resonance features either (see supplementary part Figure S3); thus, their EM performance is suppressed.

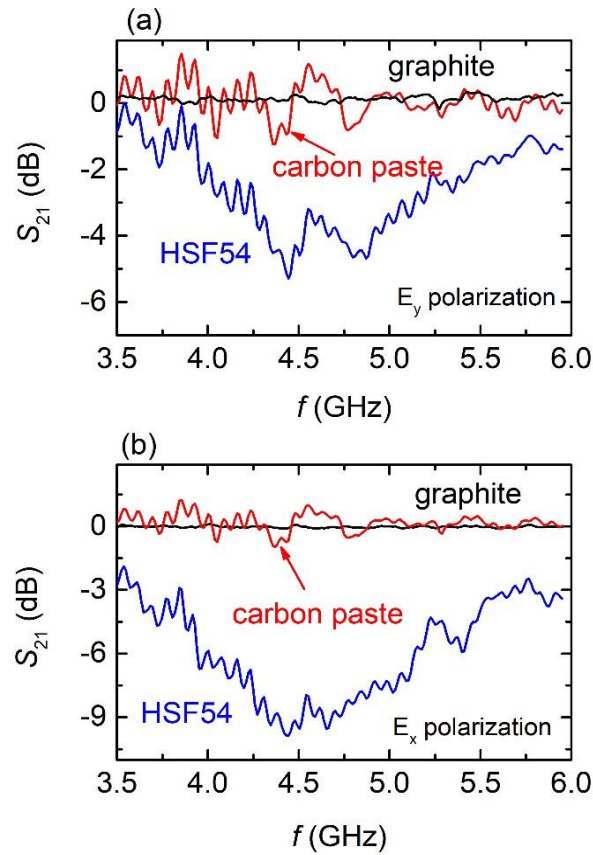

**Figure 8.** (**a**) $S_{21}$ spectra of all cut–wire MSs printed on wood surface. The spectra were obtained in the $E_y$ polarization. (**b**) $S_{21}$ spectra for wood printed cut–wire MSs in the $E_x$ polarization.

## 4. Summary and Conclusions

In the current study, spray-printed metasurfaces were developed on materials used in building construction, such as plasterboard and wood. Two different kinds of metasurfaces were deposited: cut-wire MSs, appropriate for energy harvesting, and Split Ring Resonators (SRRs), suitable for sensing applications.

The conventional spray process was employed to adapt such MSs to both plasterboard and wood surfaces. Both commercially available and homemade, water-based solutions were used as painting precursors. All of them include carbon nanoinclusions. A commercially available airbrush was used so that the development of the MSs onto the construction materials resembles the painting of the interior walls of a building. In order to achieve the desired shape and dimensions of the MSs, 3D printed shadow masks were used.

The spray-deposited MSs were characterized regarding their morphological, structural, conducting and electromagnetic properties. All of them possess dimensions analogous to the corresponding shadow masks, denoting an efficient printing quality, and thus the sufficiency of the growth method used. Furthermore, the shape of the MSs is moderate, with well-defined lines of uniform width. However, the corners are not as sharp as they need to be, while MS gaps are not well-shaped in some cases. Regarding their structural properties, all MSs show XRD, FT-IR and Raman spectra, which are typical for samples including graphite/carbon black nanoinclusions. As far as their resistance, in all cases the HSF54 MSs show the lowest resistance values among all precursors used.

The electromagnetic performance of all MSs printed on both plasterboard and wood was investigated. Both graphite and carbon paste MSs show poor EM behavior; thus, it seems that they cannot be promising candidates for EM applications. In contrast, the HSF54 MSs exhibited efficient EM behavior regardless of the surface on which they were developed. Specifically, the HSF54 cut-wire MSs qualitatively resemble the EM behavior of corresponding PCB-printed MSs. On the other hand, the HSF54-made SRRs also show EM characteristics similar to the corresponding PCB-printed SRRs. Therefore, the HSF54 printed MSs could potentially be used for energy harvesting, as well as sensing applications in the microwave regime. In conclusion, spray deposition of MSs in large areas such as building walls is accessible, while their potential use in microwave applications for purposes of buildings' energy autonomation is realized.